\documentclass{PoS}

\title{Pion and photon production in heavy ion collisions}

\ShortTitle{Pion and photon production in heavy ion collisions}

\author{\speaker{Gabor David}\thanks{This work has been supported by
        the Office of Science of the United States Department of Energy}\\
        BNL, USA\\
        E-mail: \email{david@bnl.gov}}

\abstract{
  Measurement of neutral pions and direct photons are closely
  connected experimentally, on the other hand they probe quite
  different aspects of relativistic heavy ion collisions.  In this
  short review of the $\pi^0$ results from the PHENIX experiment at
  RHIC our focus is on the $\phi$-integrated nuclear modification
  factor, its energy and system size dependence, and the impact of
  these results on parton energy loss models.  We also discuss the
  current status of high $p_T$ and thermal direct photon
  measurements both in $p$+$p$ and Au+Au collisions.  Recognizing the
  advantages of measuring not only the ``signal'', but also all the
  ``references'' needed for proper interpretation in the same
  experiments (with same or similar systematics) we argue that RHIC
  should regularly include $d$+A and even $d$+$d$ collisions into its 
  system size and  energy scan.
          }

\FullConference{High-pT Physics at LHC - Tokaj'08\\
		 March 16 - 19 2008\\
		 Tokaj, Hungary}

\begin{document}

\section{Introduction}
\label{sec:intro}

\begin{figure}
\begin{centering}
\includegraphics[width=.6\textwidth]{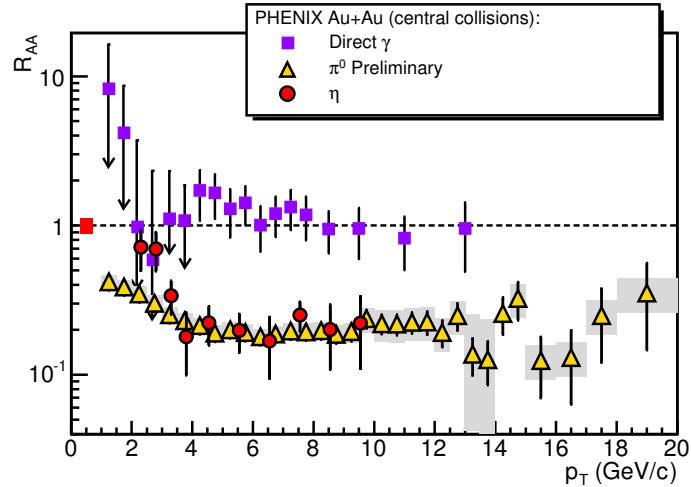}
\caption{``Historic'' nuclear modification factors $R_{AA}$ {\it vs} 
  $p_T$ in  200GeV/A Au+Au  collisions.  Squares: direct photons (no 
  apparent  suppression~\cite{ppg042}, published 2005).  
  Triangles and circles: preliminary results on $\pi^0$ and $\eta$,
  also from 2005. }
\label{fig:gpe_RAA}
\end{centering}
\end{figure}

One of the {\it predicted} signals of the formation of a quark-gluon
plasma (QGP) in relativistic heavy ion collisions was the
in-medium energy loss of hard scattered partons, leading to ``jet
quenching'' and manifesting itself in a drastic reduction of high
$p_T$ hadron production in relativistic heavy ion collisions, as
compared to the production rate in $p$+$p$ scaled by the nuclear thickness
function $T_{AA}$ to account for the increased probability of hard 
scattering.  The effect has been characterized by the 
{\it nuclear modification  factor} $R_{AA}$ defined as

$$
R_{AA}^{\alpha}(p_T) = \frac{1}{N_{evt}}
\frac{d^2N_{AA}^{\alpha}/dp_Tdy}
     {\left\langle T_{AA} \right\rangle d^2\sigma_{pp}^{\alpha}/dp_Tdy}
$$

for an arbitrary particle type $\alpha$.
In the absence of nuclear effects $R_{AA}=1$, otherwise it is enhanced
($R_{AA}>1$, like in the ``Cronin-effect'') or suppressed ($R_{AA}<1$
like in jet quenching.)  The first major discovery at RHIC was that in
central Au+Au collisions hadrons were strongly 
suppressed~\cite{ppg003,ppg014} 
(see Fig.~\ref{fig:gpe_RAA}), in accordance with the predicted energy
loss of partons in the hot, dense medium.  Lack of suppression in
d+Au collisions~\cite{ppg028} (where we assume no such medium is formed) 
supported the conclusion that the jet suppression was indeed a final
state effect, as opposed to some drastic changes in the initial state
(parton distribution functions).  
The first observation that direct 
photons were not suppressed, not even in the most central Au+Au
collisions~\cite{ppg042} made this first, qualitative picture
coherent, by showing that the crucial step in forming $R_{AA}$ -
scaling $p$+$p$ cross sections by $\left\langle T_{AA} \right\rangle$ to
obtain the expected Au+Au yields - makes sense: photons will not
interact with the medium ($\alpha_{em} << \alpha_s$), so if the
$\left\langle T_{AA} \right\rangle$ scaling is the right concept, the
direct photon $R_{AA}$ should be unity - and indeed, it is 
(see Fig~\ref{fig:gpe_RAA}). 

While very useful and providing many early insights, causing a flurry
of theoretical activities rarely seen before, the above picture
turned out to be only rudimentary.  As RHIC experiments moved 
``from the discovery phase to the exploration phase'' more precise data
became available, with higher statistics, extended $p_T$ ranges
and more complex physics quantities (like two- and three-particle
correlations, azimuthal anisotropies, {\it etc}.
Equally important, RHIC launched a systematic energy and colliding
species scan.  This made measuring excitation functions in the same
collider and within the same experiments possible.  The collider
environment ensures that multiplicities at midrapidity rise only
slowly with $\sqrt{s_{NN}}$, and measuring in the same experiment
means that most systematic errors will be similar - even if individual
measurements are not very precise, the evolution with energy and/or
system size can be established quite accurately.  Increased accuracy
of the data made comparisons with theories more
meaningful, put some constraints on free parameters~\cite{ppg079} and
gave the first experimental indications at what region of the phase
diagram the transition to the strongly coupled quark gluon plasma
might occur.  Almost needless to say that in the process our initial,
relatively simple picture became richer (much more complicated).

\section{Pion suppression systematics}
\label{sec:pionsuppression}

So far four aspects of the $\phi$-integrated  $R_{AA}$ systematics 
have been studied at RHIC.  First, the magnitude and shape of
$R_{AA}(p_T)$ at the highest 
available transverse momenta - namely, whether it remains constant, 
consistent with a constant fractional energy loss picture, or rises 
with increasing $p_T$, as predicted by most parton energy loss
models~\cite{ppg079}.  These studies gave the first quantitative
constraints on free parameters of models, like initial gluon density
or the transport coefficient of the PQM~\cite{PQM2007} model
$\left\langle \hat{q}=\mu^2/\lambda \right\rangle =
13.2^{+2.1}_{-3.2}$GeV$^2$/fm, where $\mu$ is the average transverse
momentum transfer to the medium per mean free path $\lambda$.
PQM - which includes radiative loss, a static medium, no initial state
multiple scattering and unmodified PDFs - predicts a slow rise 
of $R_{AA}$ with $p_T$, as do other models with very different
assumptions~\cite{ppg079}.  In light of this it is interesting, that 
the data are best described by a constant fit which in turn for the
power law spectra of high $p_T$ pions would imply a constant fractional
energy loss (constant $S_{loss}=\Delta p_T / p_T$ fractional shift 
of the spectrum).  These statements are based on data taken in 
RHIC Run-4 (2004), which still have large errors on the high $p_T$
points, results from the much higher statistics 2007 data should be
available soon.

\begin{figure}
\begin{centering}
\includegraphics[width=.6\textwidth]{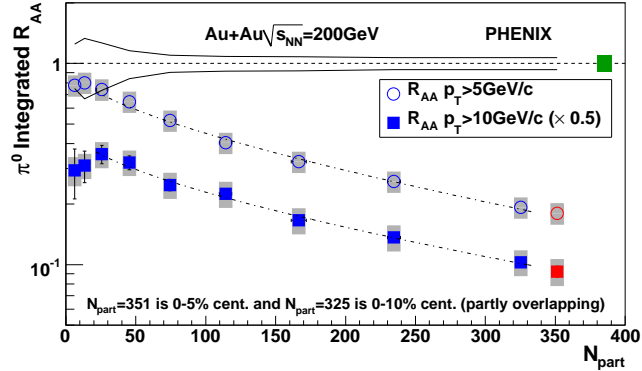}
\caption{Integrated $R_{AA}$ for $\pi^0$ as a function of collision
  centrality expressed in terms of $N_{part}$.  The last two points
  correspond to overlapping centrality bins 0-10\% and 0-5\%.  See
  text for an explanation of the fits.}
\label{fig:int_raa}
\end{centering}
\end{figure}

Second, the evolution of $R_{AA}$ with collision centrality has been 
studied.  One very interesting and non-trivial observation was that the
shape of the high $p_T$ $\pi^0$ spectra is virtually unchanged from
$p$+$p$ to the most central Au+Au collisions.  PHENIX found~\cite{ppg080}
that the exponent of the power law function fitting the high $p_T$
part of the spectra ($\propto p_T^n$) varies from 
$n=-8.22\pm0.09$ in $p$+$p$ to $n=-8.00\pm0.012$ in the most central
(0-5\%) Au+Au collisions.  This combined with the fact that $R_{AA}$
in the most central collisions is essentially flat (constant) means
that $R_{AA}$ is constant at {\it all} centralities to the best of our
current knowledge which in turn means that $R_{AA}$ integrated above a
certain $p_T$ is a useful quantity to characterize the onset and
evolution of jet quenching.  This is shown on Fig.~\ref{fig:int_raa}.  
Fitting the points with a function motivated by the constant
fractional energy loss ({\it i.e} the energy loss increases with $p_T$)
$R_{AA} = (1 - S_0 N^a_{part})^{n-2}$ the exponent $a$ is found to be
$a=0.58\pm0.07$ for $p_T>5$GeV/c and $a=0.56\pm0.10$ for $p_T>10$GeV/c 
which is in reasonable (albeit not perfect) agreement with the 
$a \approx 2/3$ predictions of the PQM~\cite{PQM2007} and GLV~\cite{GLV2000}
 models.  Note that in the ``constant fractional
energy loss'' picture $R_{AA}$ depends on the slope (exponent $n$) of
the power-law $p_T$ spectrum which in turn depends on the per-nucleon
collision energy $\sqrt{s_{NN}}$.

\begin{figure}
\begin{centering}
\includegraphics[width=.6\textwidth]{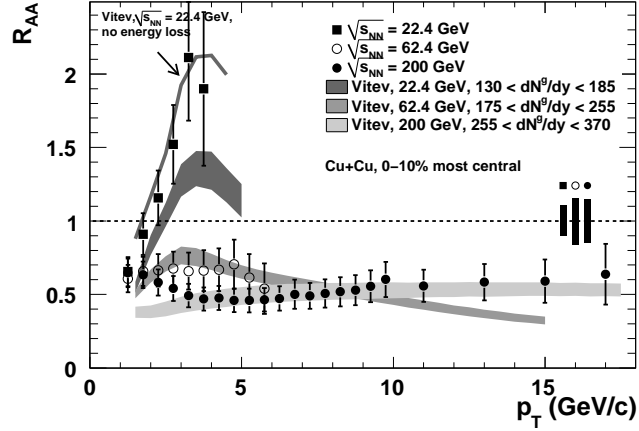}
\caption{Nuclear modification factors for the most central Cu+Cu
  collisions~\cite{ppg084} for $\sqrt{s_{NN}}=$22.4,62.4 and 200GeV
  compared to theoretical calculations~\cite{vitev2006}.}
\label{fig:raa_CuCu}
\end{centering}
\end{figure}

Therefore, the third issue is what can be learned from the energy
dependence of $R_{AA}$, and the first answer is given on
Fig.~\ref{fig:raa_CuCu} which shows the nuclear modification factors
measured at RHIC in the most central Cu+Cu collisions at 
$\sqrt{s_{NN}}=$22.4,62.4 and 200GeV.  Instead of suppression the
lowest energy data show a substantial {\it enhancement} of $R_{AA}$ in
line with the well-known ``Cronin-effect'' (and within errors this
enhancement is {\it independent} of centrality~\cite{ppg084}) whereas at 
$\sqrt{s_{NN}}=$62.4GeV the pions are already significantly
suppressed.  Of course it does {\it not} follow that the onset of jet
quenching - presumably due to QGP formation - happens necessarily 
between $\sqrt{s_{NN}}=$22.4 and 62.4GeV; all we can say is that the
quenching mechanism overtakes the Cronin-effect in this region.
Being a consequence of the initial $k_T$ smearing of partons the
Cronin-effect is present already in cold nuclei and could be unfolded
from the $\sqrt{s_{NN}}$-dependence of $R_{AA}$ if $p$+A ($d$+A) data were
available from RHIC in the entire available collision energy range.
In the author's opinion taking $d$+A data (along with A+A and $p$+$p$ for
reference) should be part of the planned ``energy scan'' at RHIC.
- Since pions are strongly suppressed in Cu+Cu already at
$\sqrt{s_{NN}}=$62.4GeV, it is not surprising that they are also
suppressed in Au+Au, in fact, almost as strongly as at 130 or 200GeV.  

Finally, the fourth issue is how does $R_{AA}$ depend on the
(size of the) colliding systems at any given energy.  A comparison of
$\sqrt{s_{NN}}=$200GeV central Cu+Cu collisions to mid-central Au+Au 
(such that the number of participating nucleons $N_{part}$ is the same 
in both cases albeit the geometry is not) shows similar values 
of $R_{AA}$.  It would be interesting to know which is the lightest, 
symmetric heavy ion system where jet quenching at
$\sqrt{s_{NN}}=$200GeV would disappear. 

So far we discussed only the $\phi$-integrated $R_{AA}$.  For any
given dataset neglecting azimuthal dependence results in higher
precision of the measurement itself (better statistics and no
systematic errors related to the determination of the reaction plane) 
but it also hides any information encoded in azimuthal anisotropies 
of $R_{AA}$ due to the overlap geometry of the colliding nuclei.  
Such anisotropies clearly exist: after all, some of the most
spectacular observations at RHIC are the large ``elliptic flow'' of 
hadrons and the quark number scaling thereof.  Indeed, when studying
$R_{AA}(\Delta\Phi)$ where $\Delta\Phi$ is the azimuthal angle
with respect to the reaction plane, determined event-by-event, 
we observe a rich structure~\cite{ppg054}: jet quenching varies 
substantially with the average pathlength of the parton in the medium.
Back-to-back dihadron (or more ambitiously, dijet) correlations, and
particularly the ``golden channel'', photon-hadron (photon-jet)
correlations have also been advocated - and now measured -
as tools of ``jet tomography''.  It has been argued~\cite{wang2007}
that single hadron spectra at high $p_T$ are dominated by surface
emission, therefore, $R_{AA}$ is barely sensitive to the medium deep
inside the collision volume.  On the other hand back-to-back
correlations force one of the partons to go through the medium (with
the exception of ``tangential'' emissions) thus the dihadron
suppression $I_{AA}$ is more sensitive to the input parameters of
theories like initial energy or gluon density.  This is certainly true
as long as the uncertainties of both theory and experimental data are
negligible: a multidifferential quantity always provides more
constraints.  However, theories usually have uncertainties (even if
often unstated, regrettably), and experimental data always have them
and it is hard to imagine a situation in which 
{\it using a given dataset} a multidifferential quantity could be 
measured more precisely than a single differential.  Just the
opposite.  The {\it quantitative} question then is: does the increased 
sensitivity on the theory side compensate for the inevitable loss of 
precision in the data?  The answer varies from case to case.

\section{Direct photons}
\label{sec:photons}

Direct photons - defined as those not originating from final state hadron
decays - offer unique possibilities to study heavy ion collisions.
First, there are physics mechanisms that produce direct photons at
each stage of the collision, thus they have the potential to provide
information on the entire evolution, ``history'' of the collision.
Equally important, once produced, they leave the interaction region 
mostly unaltered ($\alpha_{em}<<\alpha_s$), so the information they
carry is unbiased.  Unfortunately, the very same properties make them
challenging to measure (photons from hadron decays provide a large 
background, particularly at low $p_T$) and their ``message'' is hard 
to decypher because it is difficult (if not impossible) to deconvolute
the contributions of individual mechanisms to the observed direct
photon spectrum.

On Fig.~\ref{fig:gpe_RAA} we already have seen the first measurement
of direct photon $R_{AA}$ in central Au+Au collisions at
$\sqrt{s_{NN}}=$200GeV, showing that in the region where photons from
primordial hard scattering were expected to dominate
($p_T>$5-6GeV/$c$) $R_{AA}$ is indeed consistent with unity.  This was
consistent with the picture that both the PDFs and the production
cross section are unchanged, the only difference between $p$+$p$ and
Au+Au as far as hard scattering is concerned is the increased
probability of an otherwise rare process; this increased
probability is properly described by a straightforward geometric
overlap integral (the nuclear thickness function $T_{AA}$), and the
photons produced escape the collision region unaltered even if a
dense, strongly interacting medium is formed which quenches parton
jets.  Although we have higher quality data now (in an extended range)
and our understanding of the processes generating photons is much more
nuanced, we should emphasize that at a very basic level, in first
approximation, qualitatively the above picture is still correct. 

What are the currently known/assumed sources of direct photons in
$p$+$p$ and Au+Au collisions?
At leading order two processes generate photons, the 
quark-quark annihilation ($q\bar{q} \rightarrow g\gamma$), which is 
suppressed at high $p_T$ due to the lack of valence antiquarks, and 
the dominant quark-gluon Compton-scattering ($qg \rightarrow q\gamma$).  
Note that in the latter the photon balances the momentum of the
original quark (modulo the initial $k_T$ smearing of the partons) 
and it comes out freely from the medium, whereas the quark will lose a
significant fraction of its energy. Therefore, in back-to-back
photon-jet correlations the photon  ``calibrates'' the energy scale of 
the jet, which is important in ``jet tomography'' as well as in
studying possible modifications of the fragmentation function in the
presence of a medium.  The probe itself is very clean and well
understood - unfortunately the rates are quite low, due to the 
$\alpha_{em}$ coupling.

\begin{figure}
\begin{centering}
\includegraphics[width=.6\textwidth]{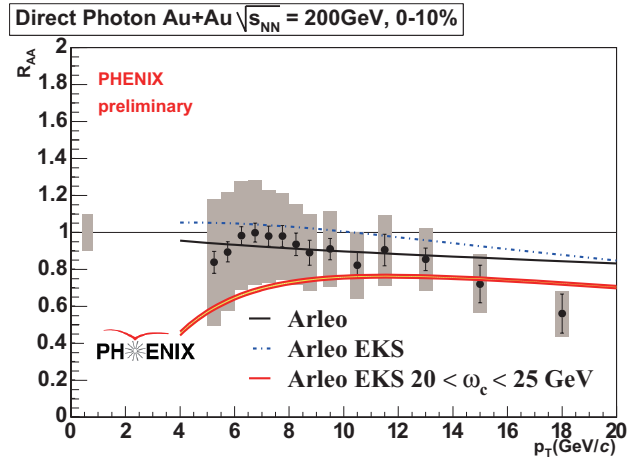}
\caption{Direct photon $R_{AA}$ in central Au+Au collisions.  The
  denominator is a fit to the PHENIX preliminary Run-5 $p$+$p$ data.
  Theoretical curves are LO calculations from~\cite{arleo2006}.  The
  solid line shows the pure isospin effect, the dash-dotted line is
  the combined effect of isospin and (anti)shadowing (EKS), finally
  the band at the botoom combines isospin and antishadowing with
  photon quenching.  }
\label{fig:raa_qm06}
\end{centering}
\end{figure}

At higher order fragmentation (or Bremsstrahlung) photons contribute
both in $p$+$p$ and heavy ion collisions.  In principle Compton and
fragmentation photons are distinguishable (at least in $p$+$p$) because
fragmentation photons should have enhanced, jet-like activity in a
small cone around them
(tracks or energy deposit that exceeds the level
justified by the underlying event),
whereas LO Compton photons should have none
in their neighborhood: they are
``isolated''.  There are predictions that at high $p_T$ fragmentation 
photons will be a larger fraction of all direct photons in A+A
collisions than in $p$+$p$.  This is another reason why is it so
important to understand $p$+$p$ processes as well as possible: they
will serve as reference when we try to find new phenomena (new sources
of direct photons) in A+A.  First direct photon cross-sections in
$\sqrt{s}=$200GeV $p$+$p$ collisions at RHIC were published
in~\cite{ppg049} followed by~\cite{ppg060} with extended $p_T$ range,
small systematic errors and detailed comparison to NLO pQCD
calculations.  Above $p_T>5$GeV/$c$ where the systematic errors of
theory and experiment are comparable, the data are well described by
the theory, including the fraction of isolated photons

Is $p$+$p$ the right reference at all?  For pions it certainly is due to
the isospin symmetry of the protons and neutrons of the nucleus.  The
same is not true for direct photons: due to the $\alpha_{em}$ coupling
the photon cross section is proportional to the sum of the squared
quark charges $\Sigma e^2_q$ which is different for $p$ and $n$.  This
``isospin effect'' causes a trivial shift of the direct photon
$R_{AA}$~\cite{arleo2006}, 
depending on the nucleus and to some small extent on
centrality; in the absence of any other medium or nuclear effect
$R_{AA}$ should go about 10-15\% below unity in Au+Au collisions 
(see the solid curve on Fig.~\ref{fig:raa_qm06}, which shows the
latest, still preliminary PHENIX results for photon $R_{AA}$ in the
most central Au+Au collisions).  Note that this
trivial effect may either mask or enhance other mechanisms affecting
$R_{AA}$.  The clean way to generate direct photon $R_{AA}$ would be
to compare to a proper mixture of photon cross sections in 
$p$+$p$, $p$+$n$ and $n$+$n$ collisions, 
which could be readily available at RHIC from 
$d+d$ collisions, since these three types of collisions could easily
be tagged.

\begin{figure}
\begin{centering}
\includegraphics[width=.6\textwidth]{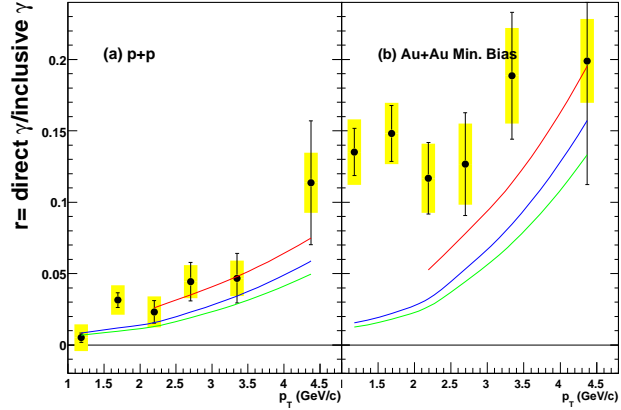}
\caption{The fraction of the direct photon component as
  a function of $p_T$ in (a) $p$+$p$ and (b) Au+Au (min. bias).
  The error bars and the error band represent the statistical and
  systematic uncertainties, respectively. The curves are
  from a NLO pQCD calculation.}
\label{fig:thermal}
\end{centering}
\end{figure}

A recent overview of the possible (additional) photon production
mechanisms in heavy ion collisions is given in~\cite{turbide2008}.  
At medium $p_T$ (corresponding to the $0.1<x<0.2$ region)
anti-shadowing should enhance the photon yield.  The large parton
energy loss (deduced from $\pi^0$ suppression) means that
fragmentation photons could be suppressed in the entire known $p_T$
range, {\it unless} the fragmentation functions themselves change in
the presence of a medium (which is hard, but not impossible to measure
in Au+Au).  On the other hand the jet-photon conversion mechanism may
increase high $p_T$ photon production.  In this process a hard
scattered quark interacts with a gluon or antiquark of the medium and
the outcoming photon carries essentially the full momentum of the
original parton (hence the name).  Disentangling the individual
contributions of these processes will not be easy, although we should
point out that some mechanisms produce isolated, others give
non-isolated photons, and their azimuthal
asymmetries are also different (for instance jet-photon conversion
should produce isolated photons with a negative ``elliptic flow''
$v_2<0$ which is quite unique since all other particles tested so far 
at RHIC exhibited $v_2>0$).

Low $p_T$ real photons (thermal region) are very hard to measure due 
to the overwhelming hadron decay background, however, ``quasi-real''
photons are accessible as low mass $e^+e^-$ pairs.  Using this
technique PHENIX measured the fraction of direct photons in the total
inclusive photon spectrum~\cite{thermal_photons} 
(see Fig.~\ref{fig:thermal}) and found that while in $p$+$p$ the
results are consistent with NLO pQCD calculations (there are no
unexpected sources), in Au+Au there is a statistically significant
10\% excess at low $p_T$.  

In summary, after the early, exciting discoveries RHIC experiments 
entered the era of exploration.  Data became more precise enabling
detailed comparisons to theories and setting meaningful constraints on
their free parameters.  Taking advantage of the flexibility of the
accelerator a deliberate study of energy- and system-size dependence
of the new phenomena is underway.  The lively cooperation of theorists
and experimentalists (strengthened by some new initiatives) will
certainly lead to our much deeper understanding of the strongly
coupled quark-gluon plasma.

\end{document}